\documentclass[twocolumn,showpacs,aps,prb]{revtex4}
\usepackage{graphicx}

\begin{document}



\title{Efficient switching of Rashba spin splitting in wide modulation-doped quantum wells}

\author{D.M. Gvozdi\'{c}}
\affiliation{Faculty of Electrical Engineering, University of
Belgrade, P.O. Box 35-54, Belgrade, Serbia}
\author{U. Ekenberg}
\altaffiliation{Corresponding author. Electronic address:
ekenberg@imit.kth.se} \affiliation{School of Information and
Communication Technology, Royal Institute of Technology, SE-16440
Kista, Sweden}

\begin{abstract}

We demonstrate that the size of the electric-field-induced Rashba
spin splitting in an 80 nm wide modulation-doped InGaSb quantum well
can depend strongly on the spatial variation of the electric field.
In a slightly asymmetric quantum well it can be an order of
magnitude stronger than for the average uniform electric field. For
even smaller asymmetry spin subbands can have wave functions and/or
expectation values of the spin direction that are completely changed
as the in-plane wave vector varies. The Dresselhaus effect can give
an anticrossing at which the spin rapidly flips.

\end{abstract}


\maketitle

\section{Introduction}

There is presently a strong interest in spin-related phenomena in
semiconductors and the prospects of utilizing the spin rather than
the charge of the electron for devices have given rise to a new
research area called spintronics. \cite{Zutic} One important
mechanism that can be used in spintronic devices is called the
Rashba effect. \cite{Bychkov} An applied electric field is seen in
the frame of a moving electron as having a magnetic field component
and yields a spin splitting even in the absence of magnetic field or
magnetic ions. \cite{Winklerbook} The Rashba effect is the mechanism
behind the Datta-Das spin field effect transistor \cite{Datta} which
is perhaps the most well-known spintronic device. The spin-orbit
coupling in a quantum well gives a subband splitting that is usually
described in the Rashba model by \cite{Bychkov,Silva}

\begin{equation}
\Delta E = \alpha k_{\|} = \frac{\hbar^2 \Delta (2 E_g + \Delta)}{2
m E_g (E_g + \Delta) ( 3 E_g + 2\Delta)} \, \varepsilon k_{\|}
\end{equation}

where $\alpha$ is commonly called the Rashba coefficient, $k_{\|}$
is the in-plane wave vector and $\varepsilon$ is the electric field.
For the expression for $\alpha$ taken from Ref. \onlinecite{Silva}
we insert the parameters of the well material.

The Rashba coefficient is related to the electric field
perpendicular to the quantum well but so far little attention has
been paid to the influence of the spatial variation of the electric
field. We here find that under certain circumstances insertion of
the different kinds of averages, e.g. the expectation value of the
electric field, gives incorrect results. In particular we show how
modulation-doping can give a strong Rashba effect with an applied
field being an order of magnitude smaller than in the case of
uniform doping and that it also can give rise to interesting
anticrossing phenomena.

\section{Theory}

We have gone beyond the Rashba model and performed self-consistent
subband structure calculations in the Hartree approximation in a
multi-band $\textbf{k} \cdot \textbf{p}$ envelope function approach.
The interaction between the conduction band, heavy-hole band,
light-hole band and split-off band is included exactly in an $8
\times 8$ matrix and the contributions from the remote bands are
included in perturbation theory. \cite{Winklerbook,Cohen} We include
terms due to the asymmetry of the zincblende lattice (Dresselhaus
effect \cite{Dresselhaus}) and add the macroscopic potential along
the diagonal of the matrix. This approach simultaneously gives
accurate descriptions of the electron and hole subbands. We have
here considered an 80 nm wide In$_{0.74}$Ga$_{0.26}$Sb quantum well
(QW) surrounded by In$_{0.7}$Al$_{0.3}$Sb barriers. In this way we
essentially retain the strong spin-orbit coupling of InSb and get
lattice-matched well and barrier materials with a suitable
conduction band offset.

\begin{figure}[h]
\includegraphics{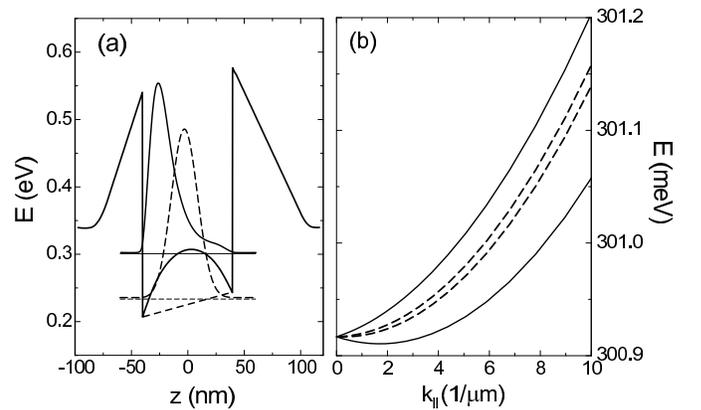}
\caption{(a) Potential and squared wave function and (b) subband
dispersion for the lowest subband pair in an 80 nm InGaSb quantum
well. The quantum well bias (potential difference between the
interfaces) is 36 mV. Dashed lines: uniform electric field, solid
lines: modulation-doped quantum well with an electron density of $6
\cdot 10^{11}$ cm$^{-2}$.\label{enhanced}}
\end{figure}

\section{Modulation-doping vs. uniform electric field}

To illustrate one important effect of ours we compare the situation
in a modulation-doped quantum well (MDQW) with that in a QW with a
uniform electric field. The potential difference between the
interfaces (below denoted quantum well bias, QWB) is the same in
both the cases, 36 meV. We here take the wave vector to be in the
[10] direction in the two-dimensional Brillouin zone. In this
direction the Rashba effect dominates over the Dresselhaus effect.
The potentials, squared wave functions and spin-split ground state
subbands are shown in Fig.~\ref{enhanced}. In the modulation-doped
case the carrier density was taken to be $6 \cdot 10^{11}$
cm$^{-2}$. We then have two weakly interacting electron gases in the
interface regions.

It is seen that the spin splitting is an order of magnitude larger
in the modulation-doped case. Relative to a symmetric QW without
Rashba splitting we present a modified mechanism to apply a moderate
QWB and take advantage of the much stronger built-in electric field
to obtain a substantial Rashba splitting. The reason for this effect
is seen from the wave functions. In a symmetric QW the wave
functions of the two lowest subbands would be symmetric and
antisymmetric, respectively, and thus spread over the entire QW. But
for sufficiently large asymmetry each wave function becomes
localized to one of the interface regions. There the electric field
is quite strong and it is this local field, not any average field,
that determines the size of the spin splitting, in contrast to
common belief so far.

\section{Wave function dependence on in-plane wave vector}

Interesting things happen if we consider a MDQW with very small QWB,
1.7 meV. This is comparable to the energy separation at $k_{\|} = 0$
between the lowest two subbands, $E_{21} = 1.4$ meV. This leads to
interesting anticrossing phenomena and the influence of the next
lowest subband must be seriously considered. We have found that
anticrossings can be influenced strongly by the Dresselhaus effect
which is stronger when $k_{\|}$ is in the [11] direction. From now
on we consider $k_{\|}$ in this direction.

\begin{figure}[h]
\includegraphics{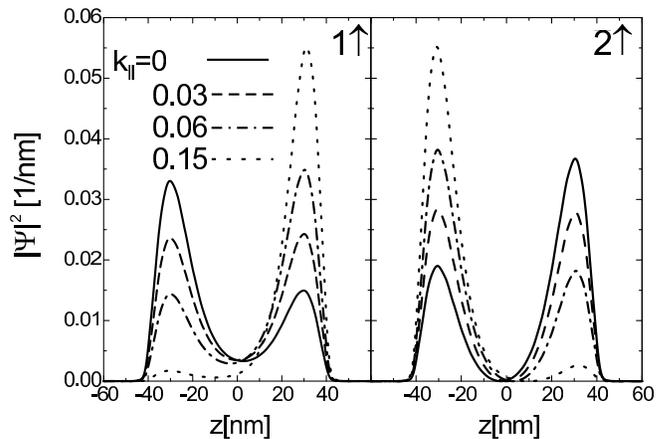}
\caption{Squared wave functions for four k$_{\|}$-values in units of
nm$^{-1}$ in an 80 nm InGaSb quantum well. The quantum well bias is
1.7 mV. Left: upper component of the lowest subband pair. Right:
lower component of next lowest subband pair.\label{psi}}
\end{figure}

The wave function at $k_{\|}=0$ for the next lowest subband is
mainly localized at the right interface region where the electric
field is reversed (cf. Fig. 1a) and therefore we have the opposite
order between "spin-up" and "spin-down" subbands. In order of
increasing energy at small $k_{\|}$ it is therefore appropriate to
label the lowest four spin subbands $1\downarrow, \, 1\uparrow, \,
2\uparrow$ and $2\downarrow$. For such a small asymmetry the wave
functions at $k_{\|} = 0$ also have a non-negligible amplitude in
the other interface region (Fig.~\ref{psi}). When we increase
$k_{\|}$ the wave functions of two adjacent spin subbands
($1\uparrow$ and $2\uparrow$) move towards the opposite interface
region, which is rather unexpected at first sight. Near $k_{\|} =
0.03$ nm$^{-1}$ it is seen that the squared wave functions have an
even distribution between the two interface regions and then the
energy separation also has a local minimum. The other two wave
functions (1$\downarrow$ and 2$\downarrow$), on the other hand,
become more strongly localized to one interface region (not shown).

\begin{figure}[h]
\includegraphics{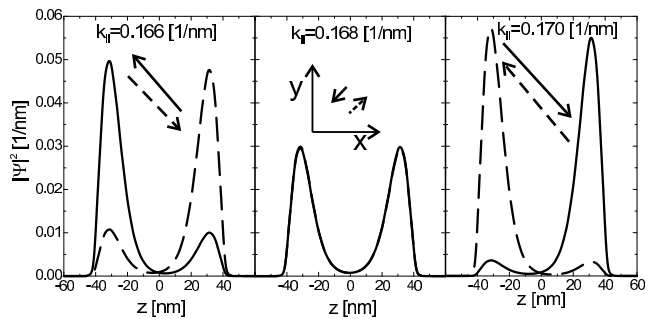}
\caption{Squared wave functions for three successive k$_{\|}$-values
(left to right). Same potential as in Fig.2. In the middle figure
the difference between the curves is too small to be visible on this
scale. The projections in the $xy$-plane of the expectation value of
the spin vector are shown for each spin subband in the insets. Solid
lines: lowest spin subband, dashed lines: next lowest spin subband.
\label{spin}}
\end{figure}

Another type of anticrossing takes place between the spin subbands
$1\downarrow$ and $1\uparrow$ around $k_{\|} = 0.168$ nm$^{-1}$
(Fig.~\ref{spin}). It is seen that in a narrow range of
$k_{\|}-$values, 0.166 to 0.17 nm$^{-1}$, the wave functions are
interchanged and, simultaneously, the expectation value of the spin
\cite{Winklerbook} for a given spin subband is flipped.

\section{Analysis of anticrossing phenomena}

The interchange of properties is typical for anticrossing of
subbands. For uncoupled spin subbands the Rashba model (Eq. (1))
predicts a linear increase of the energy splitting with $k_{\|}$ and
it is clear that eventually it would exceed $E_{21}$ and the spin
subbands $1\uparrow$ and $2\uparrow$ would cross. In our multi-band
approach an anticrossing takes place around $k_{\|} = 0.03$
nm$^{-1}$ between these spin subbands instead. This anticrossing
takes place over a rather wide range of $k_{\|}$-values. Since these
subbands have parallel spins no significant modification of the spin
expectation values takes place.

The first anticrossing described above makes the character the
second anticrossing possible. Between the anticrossings the next
lowest spin subband ($1\uparrow$) has the opposite wave function
localization and spin direction compared to the lowest spin subband
$1\downarrow$. Fig.~\ref{spin} displays a different mechanism
compared to the gradual spin precession utilized in the Datta-Das
spin transistor \cite{Datta}. The weak interaction between these two
spin subbands makes it possible to reach an energy separation of
only 0.4 meV and have such a rapid interchange of properties as
$k_{\|}$ increases. Inclusion of the Dresselhaus effect is essential
to get this behavior. Although the spin flip here occurs as $k_{\|}$
increases it should be possible to design a structure where the spin
direction at the Fermi level of a spin subband is reversed when the
bias is changed slightly.

It is clear that the wave vector range during which the anticrossing
takes place strongly depends on the spin directions of the
anticrossing spin subbands. The anticrossing can be conveniently
controlled by well width, spacer layer widths, carrier density and
applied bias.

One may argue that our results could be explained within the common
Rashba model provided that we insert into Eq. (1) the expectation
value of the electric field, which can be expected to be enhanced by
the localization of the wave function. This procedure would be
well-defined if both the spin subbands had the same expectation
value of the electric field. However, for small electric fields we
find that the expectation values averaged over the filled states (or
evaluated at the Fermi energy) can be quite different for the
different spin subbands as a result of the strong wave vector
dependence of the wave functions.

\section{Device aspects}

The strong enhancement of the Rashba splitting described in Fig. 1
due to modulation-doping can be expected to have important
implications for spintronic devices like the spin transistor
proposed by Datta and Das. \cite{Datta} For its performance it is
essential that one can achieve a large wave vector splitting $\Delta
k$ of a spin-split subband with a small bias. Utilizing the built-in
electric field one can achieve a given $\Delta k$ with a QWB that is
an order of magnitude smaller than with a uniform electric field. We
have previously \cite{Gvozdic} approximated the switch energy for
n-type and p-type spin transistors by $C V^2$, where $C$ is the
capacitance of a QW structure surrounded by two gates and $V$ is the
applied bias between them. We then concluded that n-type spin
transistors with the original design would have problems to become
competitive with conventional transistors unless fundamentally new
ideas were presented. If we only consider the lowest spin subband
pair and follow the approach of Ref. \onlinecite{Gvozdic} we obtain
a switch energy of 0.4 aJ in the modulation-doped case and 35 aJ in
a spin transistor with the same length and uniform electric field.
The former figure compares very well with present state-of-the-art
transistors \cite{ITRS} where 3 aJ has been projected.

However, a complication with our design is that the second subband
pair with the opposite sign of $\Delta k$ and spin precession
direction is also filled. This does not prevent the possibility that
the spins at the interfaces can have made a precession by the angle
$\pi$ but in opposite directions on the arrival to the drain.
Furthermore the matter is complicated by the $k_{\|}$-dependent wave
functions and the redistribution of carriers in the QW. Still it is
clear that interesting possibilities occurring from the controllable
properties open up for the design of modified spin transistors,
especially if one manages to contact the electron gases in the
interface regions separately. Such considerations will be presented
elsewhere.

\section{Discussion}

The effects described here also apply to p-type QWs. However, we
have recently demonstrated \cite{GvozdicEPL} that for p-type spin
transistors the optimal choice is quite a small electric field
($\sim$ 2 - 5 kV/cm) which is remarkably efficient to create a huge
Rashba splitting $\Delta k$.

We have implicitly assumed coherence of the wave function across the
80 nm QW with a high and broad barrier in the middle. Whether this
coherence actually prevails should depend on the sample quality.
This system with our predicted effects seems ideal for further
studies of this fundamental problem.

\section{Summary}

In conclusion we have demonstrated that the non-uniform electric
field in wide modulation-doped quantum wells gives interesting and
useful effects. One can use a bias corresponding to a moderate
average electric field and still get a Rashba splitting typically
enhanced by an order of magnitude due to the built-in local electric
field in the interface region. The switching mechanism is based on
localization of the wave function to one interface region with a
barely sufficient bias. For very small bias the wave functions and
spin directions can become strongly dependent on the in-plane wave
vector. At anticrossing of spin subbands the wave function moves
towards the opposite interface as $k_{\|}$ increases and sometimes
the spin is also flipped. The device prospects are promising but
require further analysis.

\section{Acknowledgment}

D.M.G. wishes to acknowledge the hospitality of the School of
Information and Communication Technology of the Royal Institute of
Technology where much of this work was carried out. Funding has been
received from the Swedish Research Council and the Serbian Ministry
of Science and Environmental Protection.

\end{document}